# Priority Questions for Jupiter System Science in the 2020s and Opportunities for Europa Clipper

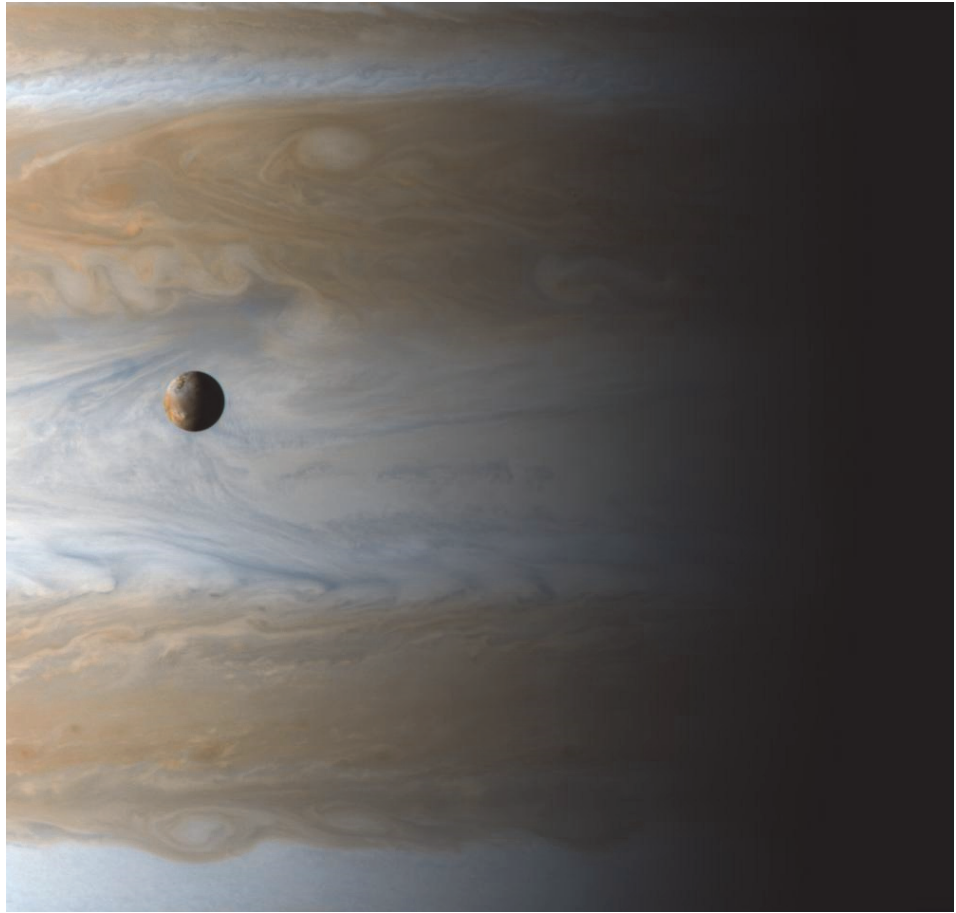

Io is seen floating above Jupiter's cloud deck during Cassini's distant flyby of Jupiter on January 1, 2001. What will we see 30 years later on January 1, 2031?

Kunio M. Sayanagi (Hampton University, kunio.sayanagi@hamptonu.edu)

Tracy Becker (SwRI)
Shawn Brooks (JPL/Caltech)
Shawn Brueshaber (Western Michigan U.)
Emma Dahl (New Mexico State University)
Imke de Pater (UC Berkeley)
Robert Ebert (SwRI)
Maryame El Moutamid (Cornell University)
Leigh Fletcher (University of Leicester)
Kandi Jessup (SwRI)
Alfred McEwen (University of Arizona)
Philippa M. Molyneux (SwRI)
Luke Moore (Boston University)
Julianne Moses (SSI)
Quentin Nénon (UC Berkeley)
Glenn Orton (JPL/Calteh)
Christopher Paranicas (JHU/APL)
Mark Showalter (SETI Institute)
Linda Spilker (JPL/Caltech)
Matt Tiscareno (SETI Institute)
Joseph Westlake (JHU/APL)
Michael H. Wong (UC Berkeley/SETI Inst.)
Cindy Young (NASA LaRC)

## 1. Overview

This whitepaper identifies important science questions that can be answered through exploration of the Jupiter System. With multiple potentially habitable large moons orbiting a central object with a composition like that of the Sun, the Jupiter System is analogous to a star system. Studying the Jupiter System will address fundamental questions about how planetary systems work, how planets form/evolve, and how life can emerge under different conditions.

A particular emphasis of this white paper is on the scientific questions that can be addressed by the Europa Clipper mission. Although the science objectives of JUICE cover the entire Jupiter System, only Europa science is allowed to drive development of Europa Clipper, which is appropriate to control costs. ***Even with the Europa-focused development, the orbiter will have considerable capabilities to study the Jupiter System, and we urge that such science be allowed after launch when expanding the scientific scope can no longer affect development costs.***

The Clipper mission can examine the Jupiter System in ways not possible from Earth through, e.g., in-situ measurements of the magnetosphere and close-range observations from a wide range of phase angles. After Jupiter-orbit insertion, the orbiter spends ~one year before entering Europa-observation orbits. As of this writing, the Clipper's orbital tour has 77 orbits around Jupiter, which includes 51, 6 and 7 flybys of Europa, Ganymede, and Callisto, respectively, implying that at least 13 of the orbits do not include flybys of the current Clipper science targets[1], during which Jupiter System Science investigations could be performed without additional instrument performance requirements so long as the mission is given sufficient resources. We suggest treating the Jupiter System Science as a target of opportunity and opening a participating scientist program shortly after launch to develop Jupiter System Science investigations.

Since Clipper's potentials for Europa, Ganymede and Callisto are already well-recognized[2], we advocate to add Io, Rings, Small/Irregular Satellites, Magnetosphere, and Jupiter itself to the scope of the Clipper mission. Some of these questions may be answered only partially by Clipper and JUICE, and points to the need for more future missions to the Jupiter System. We also note the importance of observations from Earth and space telescopes to place scientific results of visiting spacecraft missions in temporal and spectral contexts and enhance their scientific return.

We also recommend the decadal survey to encourage the planetary science community to foster an interdisciplinary, diverse, equitable, inclusive, and accessible environment.

## 2. Potential Synergy between Clipper and JUICE for Jupiter System Science

JUICE will be launched in 2022, orbit Jupiter starting in 2029 and offer significant opportunities for coordinated Jupiter System observations alongside Clipper. ***We advocate for coordinated observations of the Jupiter System by Clipper and JUICE.*** Both Clipper and JUICE evolved from the Europa Jupiter System Mission (EJSM) concept, which had two spacecraft (JEO by NASA and JGO by ESA) conducting both complementary and synergistic science. Clipper and JUICE offer significant potential for synergistic collaborations like those envisioned for EJSM and recognized by the two joint meetings to discuss coordinated observations by JUICE and Europa Clipper[2]. The two teams are currently discussing coordination of satellite science investigations[1] (e.g., cross-calibration of observations, sampling of satellites from multiple vantage points).

Clipper and JUICE's simultaneous presence in the Jupiter System offer significant potential for Jupiter System Science. The Jupiter System is physically large and exhibits dynamic temporal activities (e.g. volcanism on Io, meteorology and aeronomy on Jupiter, auroral emissions at Jupiter and some satellites). Thus, understanding how physical processes vary as a function of both location and time is a challenge. Jupiter System exploration to date has provided observations through a small window in time and space with insufficient overlap to offer a holistic view. Consequently, observations from different epochs have led to very different interpretations. Coordinated observations by Clipper and JUICE offer an additional value that exceeds the sum of separate studies. For example, JUICE lacks any coverage in the mid-IR like E-THEMIS, while Clipper lacks sub-mm capabilities like SWI. Simultaneous observations of various targets from different angles/illuminations or in-situ at different locations will offer comprehensive view of active phenomena, e.g. simultaneously exploring different parts of magnetosphere while also observing auroral structure to correlate the behaviors of aurora and the magnetosphere. Observations by both spacecraft are needed to measure satellite dynamics, key to understanding the coupled tidal heating and orbital evolution[3]. Io's active eruptions feed the Io plasma torus and populate the entire system; however, how eruptions on Io affect the magnetosphere remains to be fully understood because the observations are disconnected in space and time. The two missions arriving the Jupiter System at different times will also offer an opportunity to extend the temporal coverage of tracking active phenomena (e.g., storms, belt/zone variation, etc.) beyond what could be done by either mission. Finally, future Jupiter System missions that follow Clipper and JUICE will significantly benefit from their detailed exploration of additional targets in the Jupiter System.

---

[1] https://www.lpi.usra.edu/opag/meetings/feb2020/presentations/Pappalardo.pdf
[2] https://www.lpi.usra.edu/opag/meetings/feb2020/presentations/Senske.pdf

## 3. Overarching Theme of Atmospheric Science Goals

Studies to date have characterized Jupiter's zonal circulation, discrete vortices, vertical stratification, and distributions of clouds/hazes. The next focus of atmospheric investigations is the meridional overturning circulation and temporal variability. Such studies will further our understanding of atmospheric turbulence, waves, radiative forcing, internal heat flux, and moist/dry convection, all of which operate in all other atmospheres and lead to diverse conditions. Jupiter acts as a natural laboratory by offering conditions – e.g. the lack of topographic surface, fast planetary rotation, and the vast size – that enable us to disentangle these processes to understand their effects. Many of these objectives were left unaccomplished by the Galileo mission because of an antenna problem: e.g. while the Galileo Science Requirement stated 50,000 images[4] are needed over ~3 years[5], only 1886 images of Jupiter were returned. The brief Cassini flyby of Jupiter did not fully resolve the problem when it returned about 25,000 images of Jupiter. Coordinated observations by Clipper and JUICE can recover and extend the lost Galileo objectives.

### *3.1: How does Jupiter's atmosphere regulate the planet's thermal evolution?*

All planets thermally evolve by radiating their primordial heat to space. A thick atmosphere plays a crucial role because the energy must be transported by the atmosphere before being radiated to space. In Jupiter, the heat is believed to be transported primarily through convection in the troposphere and at depth, and radiation in the stratosphere and above; however, how the combined effects of cumulus convection, meridional overturning circulation, and cloud/haze distribution lead to the observed radiated heat is largely unknown[6]. With sufficient modeling and ground-based support, Clipper's mid-IR camera could map thermal emission in discrete weather systems that cannot be resolved from Earth to advance our understanding of how the atmosphere regulates the internal heat flux and governs the planet's thermal evolution.

---

[3] https://www.kiss.caltech.edu/final_reports/Tidal_Heating_final_report.pdf
[4] Galileo Science Requirements Document 625-50, Rev. C, JPL, February 12, 1982
[5] A Science Rationale for Jupiter Orbiter Probe 1981/1982 Mission, 660-26, JPL, August 1976
[6] Li et al. Journal of Geophysical Research 117, CiteID E11002

*3.2: How is the meridional circulation structured, and coupled to the zonal circulation?*
The structure of Jupiter's meridional overturning circulation remains unknown. Based on the locations of thunderstorms, latitudinal zonal wind shear, and the cloud-top temperatures, two-storied circulation cells that reverse directions at depth have been proposed[7]. The terrestrial Hadley circulation is coupled to the mid-latitude jetstreams through eddies[8]; an analogous theoretical framework is needed to complete our understanding of the three-dimensional circulation on the giant planets[9]. To guide such a theory, Clipper can measure tracers of vertical motions such as spectrally identifiable $NH_3$ ice (not observable from Earth) and distribution of lightning flashes.

*3.3: How is energy balanced in giant planet thermospheres?*
The thermospheres of giant planets are hotter than expected from the known heat sources, which include forcing from the underlying atmosphere such as atmospheric waves and redistribution through dynamical processes, and from outside the atmosphere such as solar radiation, charged particle precipitation, and auroral heating[10]. Solar and stellar UV occultations, and mapping of $H_3^+$ emissions by Europa Clipper and JUICE can serve as inputs for thermospheric circulation models to understand variability and the interactions between dynamics, chemistry, and radiation.

*3.4: How are large, intense, colorful atmospheric vortices generated and maintained?*
Jupiter harbors atmospheric vortices of various scales including the Great Red Spot, which is recently shrinking[11]. These vortices are highly dynamic and redistribute energy and other chemical tracers. The chromophores that color them also remain inconclusive[12]. Close-range imaging observations from a wide range of phase angles not accessible from Earth by future missions should track temporal evolutions of these vortices to characterize their internal flow, differentiate between candidate chromophores[13] and characterize aerosol properties. Vortex velocity fields are sensitive to the ambient deformation radius[14], so Clipper can also reveal vertical structure at a range of latitudes by measuring winds within vortices that are too small to be studied from Earth.

**4. Overarching Theme of Io Science Goals**
Io is believed to be the main source of plasma in Jupiter's magnetosphere; however, how the volcanic plumes become ionized and redistributed, and how much material they supply to the Ocean Worlds are not fully understood. Thus, understanding Io's impact on its surroundings is critical to interpreting surface composition, interior, and the origins of the other Galilean Moons.

*4.1: How does Io supply the plasma in Jupiter's magnetosphere?*
Io's volcanoes supply 0.7-3 ton/s of neutral material[15]. Neutrals leave Io's atmosphere through, e.g., Coulomb collisions with the corotating ions in Jupiter's magnetosphere. The neutrals near Io and its orbit become ionized[16] and flow outward in the magnetosphere. Examining outward radial transport has revealed higher densities and 2-4 faster outflow rates during Io active times[17]. The activity of one of Io's volcanoes has been shown to be linked to the moon's true anomaly[18]. Future investigations should shed light on the relationship between Io's true anomaly and its net volcanic output, the variability of plasma transport through the vast Jovian system, and the loss processes.

---

[7] Ingersoll et al., 2000. Nature 403, pp. 630-632
[8] E.g. Andrews, Holton and Leovy 1987 "Middle Atmosphere Dynamics," Academic Press.
[9] Ingersoll et al. 2017. Geophysical Research Letters 44, pp. 7676-7685
[10] Muñoz et al. 2017. In Book: Handbook of Exoplanets, Springer International Publishing AG.
[11] Simon et al. 2018. Astronomical Journal 155, article id. 151
[12] Baines et al. 2019, Icarus 330, p. 217-229.
[13] Sromovsky et al. 2017. Icarus 291, p.232-244.
[14] Shetty et al. 2007. Journal of the Atmospheric Sciences, 64, pp.4432-4444.
[15] Bagenal and Delamere, 2011. Journal of Geophysical Research, 116 (A5), CiteID A05209
[16] See full description of these processes in Thomas et al. 2004, Jupiter book pages 561-591, Cambridge University Press
[17] Yoshioka et al. 2018. Geophysical Research Letters, 45 (19), pp. 10,193-10,199
[18] de Kleer et al. 2019. Geophysical Research Letters, 46, (12), pp. 6327-6332

### 4.2: How is plasma heated in the vicinity of Io?

The situation near Io itself is not well understood. The moon itself has a complex interaction with the Jovian plasma, e.g., with an Alfven wing[19]. Plasma around Io has been studied through electromagnetics of the Io interaction, impacts on the plasma, and Io's auroral signature using data[20]. In addition, while the basic structure of the Io plasma torus and the nearly circumplanetary neutral distributions has been studied extensively, many outstanding questions remain. Much remains to be understood about the physical processes that provide the total power necessary to heat the plasma to observed energy densities[7].

### 4.3: How much of the surface material on the Galilean satellites is Iogenic?

Io is a prodigious source of $SO_2$ and produce O and S ions that dominate the plasma. Iogenic O and S may also be incorporated into surface ice, e.g. Iogenic sulfur compounds have been identified on Europa's surface[21]. Whether the non-ice material on Europa is a salt (indicating an endogenic source) or an acid (suggesting an exogenic one) remain controversial. Salts might be traces of a subsurface ocean whereas hydrated sulfuric acid on the surface says less about the interior. Future exploration should examine how Io supplies materials to the Ocean Worlds in the Jupiter System.

### 4.4: What is Io's role in forming and maintaining Jupiter's radiation belts?

Juno found a depletion of ~MeV electrons near Io orbit that was 3 Io diameters wide downstream, and a drop off of trapped energetic ion fluxes moving inward to Io's orbit[22]. Modeling of energetic charged particle source, loss, and transport suggest both effects may be due to wave-particle interactions[23], which can drive trapped particles into the planetary loss cone. In addition, at lower energies (<hundreds of keV), a significant loss of energetic charged particles may occur via charge exchange with neutrals. Consequently, energetic charged particle precipitation onto Io may be less than at the other Galilean satellites. Much remains to be understood about the loss of energetic charged particles near Io and near its orbit for each species and energy range.

## 5. Overarching Theme of Magnetospheric Science Goals

Jupiter's magnetic field is almost 20,000 times more intense than the dipole moment of Earth. Plasma coupled to this rapidly rotating magnetic field is accelerated to near Jupiter's rotation rate. Jupiter's magnetosphere accelerates and heats plasma to $>10^8$ K, two orders of magnitude hotter than the Sun's corona. Embedded within this hot plasma are potentially habitable ocean worlds. The relationship of the ocean worlds to the magnetospheric sources of energy and particles are central to understanding the fundamentals of habitability within a giant planetary system[24]. Jupiter represents a miniature solar system; the processes that drive the Jupiter System are the same as those that drive the evolution of our solar system and can be assessed at a more attainable scale.

### 5.1: What are the plasma transport mechanisms in Jupiter's magnetosphere?

One topic of particular importance at Jupiter is plasma transport. The plasma is created near Io's orbit, and loss mechanisms are needed to balance the total mass in the magnetosphere. About 1/3 of the mass may be lost from charge exchange while the rest flows outward and leaves the system[18]. The fluxtubes that release the plasma must then return the magnetic flux to the inner magnetosphere presumably via centrifugally-driven fluxtube interchange, although observational evidence is limited. Europa Clipper can examine plasma transport mechanisms in the middle magnetosphere[25]. Separating spatial from temporal variability is difficult from a single spacecraft. Clipper's presence in the Jovian system will be an opportunity to correlate its in-situ radiation, magnetic field and plasma measurements with JUICE's remote-sensing Energetic Neutral Atom and in-situ particle and field measurements. Measurements by two missions would be critical to distinguish spatial from temporal variations and quantify the scale of radial transport processes.

---

[19] Saur 2004. Journal of Geophysical Research: Space Physics 109 (A1), CiteID A01210
[20] Szalay et al. 2017. Geophysical Research Letters, 44 (14), pp. 7122-7130
[21] Carlson et al. 2009. In book Europa, page 283, University of Arizona Press.
[22] Paranicas et al. 2019. Geophysical Research Letters, 46(23), pp. 13,615-13,620
[23] Nénon et al. 2017. Journal of Geophysical Research: Space Physics 122(5), pp. 5148-5167
[24] Kivelson et al, 1997. Advances in Space Research, 20(2), pp.193-204.

*5.2: How does Jupiter's magnetosphere deliver energy/chemistry to habitable ocean worlds?*
Europa Clipper will look for particle transport throughout the Jovian system. Particles sourced at low energy from Io and Europa can be accelerated within the Jovian magnetosphere and provide additional sources of chemistry and energy to the ocean worlds[27]. As Section 3 states, Io is a major source of plasma; how energy and materials are then transported to the ocean worlds, and whether the transport is critical to their habitability remain important fundamental questions to be answered.

*5.3: How are particles accelerated in giant planet magnetospheres?*
Measurements of the radiation, magnetic field and plasma environment by Clipper could assess magnetospheric dynamics and the precursor processes for acceleration in the inner magnetosphere and seek dynamical processes that drive acceleration through the magnetosphere. Although the acceleration processes at Jupiter are not well-understood, they appear to be much more effective at driving very high energy particles. The coupling of the ionosphere to the magnetosphere is currently being investigated by Juno; its discovery of large potentials in the polar region shows that the aurora may be coupled to the radiation belt. Eventually, a mission like RBSP/Van Allen Probes is needed for Jupiter to understand the acceleration of particles within Giant Planet magnetospheres. Van Allen Probes were the first mission dedicated to understanding the processes that drive particle acceleration at the Earth, and had two spacecraft that revealed many processes including wave particle acceleration that drive much of the activity within the radiation belts. Understanding how Jupiter accelerates particles and how ubiquitous these processes are also of interest to astrophysical processes that accelerate Galactic Cosmic Rays[26].

**6. Overarching Theme of Rings Science Goals**
Jupiter's rings are one of the purest known archetypes of a dusty ring system, with applications to dust-dominated debris disks both within and beyond our solar system, yet much remains unknown about the complex interplay of gravitational and non-gravitational forces that determine the observed structure. Jupiter's rings also play the role of a detector, as comet impacts including Shoemaker-Levy 9 left an imprint on the rings that was discernible over a decade later.

*6.1: How do non-gravitational forces sculpt Jupiter's rings?*
As for all dust-dominated rings, non-gravitational forces play a significant role in the transport and evolution of ring particles at Jupiter. Although the basic mechanisms seem broadly understood[27], ongoing mysteries include the outward extension of the Gossamer rings beyond the source moons Thebe and Amalthea[28], apparent sorting of particles by size in different regions of the Main and Halo rings[29], and the mechanisms by which "Lorentz" resonances between particle orbits and Jupiter's magnetic field move particles from the Main ring into the Halo and determine the structure of the Halo[30]. All these questions are currently informed only by exceedingly sparse high-phase images and spectral cubes. Europa Clipper could improve on this considerably by taking not only images at high phase angles, but also images of Jupiter's shadow boundary as it cuts across the rings and images while the spacecraft is in Jupiter's shadow (which allows access to the highest phase angles without the danger of pointing sensitive instruments too close to the Sun), both of which methods were employed by Cassini to great effect to investigate the structure of diffuse dusty rings[31]. Near-infrared data from MISE at high phase angles can better characterize the dust size distribution, potentially elucidating how non-gravitational forces operate in the main ring.

---

[25] Mauk et al, 1997. Geophysical research letters, 24(23), pp.2949-2952.
[26] Krupp et al. 1998. Geophysical research letters, 25(8), pp.1249-1252.
[27] Burns et al. (2004). In Bagenal et al. (eds), *Jupiter*, Cambridge Univ. Press, 241–262.
[28] Showalter et al. (2011). Science, 332, 711.
[29] Brooks et al. (2004). Icarus, 170, 35–57, Throop et al. (2004). Icarus, 172, 59–77.
[30] Hamilton (1994). Icarus, 109, 221–240.

*6.2: How does fine-scale structure originate in Jupiter's rings?*

Jupiter's rings harbor an unexpected amount of fine-scale structure at a variety of spatial scales. Without self-gravity between particles, such structure should quickly disperse via Kepler shear. Yet small, longitudinally-confined clumps are found near tiny Adrastea[32], vertical corrugations in the Main ring have been explained as the imprint of material shed by Comet Shoemaker-Levy 9 prior to impacting Jupiter[36], and a partial arc of material seems to stay trapped within Amalthea's inclined orbit by a mechanism that remains unclear[33]. Observations at Saturn have guided theories for similar phenomena[35], but high-resolution long-duration observations at a range of viewing geometries are needed to understand the operating mechanisms at Jupiter.

*6.3: Where and what are the source bodies for the Main ring and Halo?*

Dusty rings require replenishment from source bodies due to limited particle lifetimes[31]. A narrow belt of dust between Metis and Adrastea may contain the source bodies for Jupiter's Main ring and Halo, but prevailing models of its source body size distribution are in conflict with a *New Horizons* non-detection of <500-meter bodies[36]. Clipper could address this mystery via high-resolution imaging to search for small moons, spectral data to constrain the belt's composition, and possibly also via Europa-UVS stellar occultations to characterize the optical depth of the population.

## 7. Overarching Theme of Jupiter's Small/Irregular Satellites Science Goals

Jupiter's small inner moons (Amalthea, Metis, Adrastea and Thebe), and other small irregular satellites provide opportunities to test theories of the origin and evolution of the Jupiter System as well as the Solar System as a whole. New spectral observations of the moons by a Jupiter orbiter would provide an improvement in our understanding of the compositions and potential origins of the different captured satellite groups; obtaining quality observations of these small, dark bodies from Earth is difficult. Key questions related to the origins and histories of the satellites, and their current interactions with the larger Jovian system, are outlined below.

*7.1: What is the origin of Amalthea, and is it shared with the other inner Jovian satellites?*

While the low eccentricities and inclinations of Jupiter's four small inner satellites are consistent with formation in their current locations, Amalthea's low density (0.86 g/cm$^3$) is of a porous icy body[34], and its broad 3μm absorption feature in its NIR reflectance spectrum suggests that the surface has hydrous minerals or organic material[35]. These observations are difficult to reconcile with the high temperatures (>800 K[36]) expected at Amalthea's current orbit during the period in which the Galilean moons formed, and imply that Amalthea either formed later than the major Jovian satellites, formed in a cooler region of the circumjovian nebula and migrated inward, or was captured from outside the Jupiter System. Whether Metis, Adrastea and Thebe share low bulk density and the 3μm absorption feature like those of Amalthea remains unknown. If they share Amalthea's density, then the high tidal forces from Jupiter provide an additional argument from dynamics against at least Metis and Adrastea having accreted in place[37]. Improved characterization of the inner moons' compositions and ages is necessary to determine if all share a common origin, providing valuable new constraints to models of the formation and evolution of the Jupiter System.

---

[31] Hedman et al. (2018). In Tiscareno and Murray (eds.), *Planetary Ring Systems*, Cambridge Univ. Press, 308–337.
[32] Showalter et al. (2007). Science, 318, 232.
[33] See also Throop et al. (2018), DPS Meeting Abstracts, 50, 104.09.
[34] Anderson, et al. 2005. Science, 308, 1291-1293.
[35] Takato, N., et al. 2004. Science 306, 2224-2227.
[36] Canup and, Ward. 2002. Astronomical Journal. 124, 3404-3423.

*7.2: From which population(s) were the irregular satellites captured?*

Jupiter's irregular satellites are presumably captured planetesimals because their orbits are inconsistent with formation in the same accretion disk as the regular satellites. Studies of such captured objects are crucial to testing and validating Solar System evolution models. For example, discovery of dynamical groups among the moons[38] suggests a history of collisional losses and explain recent models of irregular satellite capture that reproduce observed orbital parameters but predict a larger population of small (R<10 km) moons than observed[39]. Spectral observations of larger satellites indicate that the surface composition varies between dynamical groups[40], potentially as a result of different histories. A high-resolution survey of orbital and optical properties from a Jupiter orbiter is required to determine whether all irregular satellites have similar ages and originate from a single population, confirm the relationships between members of dynamical groups, and potentially allow new groups to be identified.

*7.3: How do the small moons interact with the rest of the Jovian system?*

The surfaces of the small satellites are processed by the Jovian plasma and meteoroids, leading to surface non-uniformities and ejecting dust that, in the case of the inner moons, feeds the rings. A close flyby of one or more of the moons would allow direct measurement of sputtering by charged particles, providing new information about the flow of Iogenic sulfur and oxygen through the Jovian magnetosphere. Future exploration should investigate possible plasma enhancements suggested by radio observations and determine whether Amalthea possesses a plasma torus[41]. Even in the absence of close encounters, remote sensing observations can examine features such as the large leading/trailing albedo asymmetry observed on the inner moons and test whether the brightening is caused by meteoroids modifying the leading hemispheres[42].

## 8. Need for Observations from Ground-based and Space Telescopes

We advocate for continued support for Earth-based observations of Jupiter. Observations from the ground and space telescopes are crucial to investigate events that last longer than any single mission. Visiting missions also require support by Earth-based observations to provide spatial, temporal, and multi-wavelength context for the spacecraft observations, particularly in the era of ELTs and next-generation arrays. Facilities like JWST and a potential Hubble replacement could also assist in wavelengths not accessible from the ground. A space telescope dedicated to planetary science would significantly enhance the science return of many missions like Clipper especially in providing persistent UV observations to fill gaps between visiting missions.

## 9. Fostering an Interdisciplinary, Diverse, Equitable, Inclusive, and Accessible Community

Finally, we advocate the need to continue efforts to ensure equitable and inclusive representation in the planetary exploration endeavor. Any mechanism to add Jupiter System Science to the Clipper mission should serve as an opportunity to improve representation of groups who are underrepresented and marginalized in planetary science and society as a whole.

---

[37] Tiscareno, et al. 2013. Astrophysical Journal, 765, L28.
[38] Sheppard and Jewitt, 2003. Nature 423, 261-263.
[39] Nesvorny et al. 2014. Astrophysical Journal. 784, 22.
[40] Grav and Holman 2004. Astrophysical Journal 605, L141-L144.
[41] Arkhypov and Rucker, 2007. Astronomy and Astrophysics 467, 353-358.
[42] Simonelli et al. 2000. Icarus 147, 353-365.